\documentclass[a4paper,12pt]{article}

\newcommand{\be}{\begin{equation}}
\newcommand{\ee}{\end{equation}}
\newcommand{\bea}{\begin{eqnarray}}
\newcommand{\eea}{\end{eqnarray}}
\newcommand{\ena}{\end{eqnarray}}

\usepackage{graphicx,amssymb,bm,latexsym}

\newcommand{\vs}[1]{\vspace{#1 mm}}
\newcommand{\hs}[1]{\hspace{#1 mm}}
\renewcommand{\a}{\alpha}
\renewcommand{\b}{\beta}
\renewcommand{\c}{\gamma}
\renewcommand{\d}{\delta}
\newcommand{\e}{\epsilon}

\newcommand{\pa}{\partial}
\newcommand{\nn}{\nonumber\\}
\newcommand{\p}[1]{(\ref{#1})}

\begin{document}

\begin{titlepage}

\begin{flushright}
KU-TP 007 \\
hep-th/0610015
\end{flushright}

\vskip .5in

\begin{center}

{\Large\bf Supersymmetric Intersecting Branes in Time-dependent Backgrounds}
\vskip .5in

{\large Nobuyoshi Ohta$^{a,}$\footnote{E-mail address: ohtan@phys.kindai.ac.jp}
and Kamal L. Panigrahi$^{b,}$\footnote{E-mail address: panigrahi@iitg.ernet.in}}\\
\vs{10}
$^a${\em Department of Physics, Kinki University,
Higashi-Osaka, Osaka 577-8502, Japan} \\
$^b${\em Department of Physics, Indian Institute of Technology,
Guwahati, 781039, India} \vs{10}

\vskip .2in \vspace{.3in}

\begin{abstract}
We construct a family of supersymmetric solutions in time-dependent
backgrounds in supergravity theories.
One class of the solutions are intersecting brane solutions and
another class are brane solutions in pp-wave backgrounds, and
their intersection rules are also given.
The relation to existing literature is also discussed.
An example of D1-D5 with linear null dilaton together with
its possible dual theory is briefly discussed.

\end{abstract}

\end{center}

\vfill

\end{titlepage}

\setcounter{page}{2}
\section{Introduction}

There has been much interest in time-dependent supersymmetric
solutions of supergravities in ten- and eleven-dimensional
spacetime because of their applications to cosmology and to our
understanding of space-like
singularities~\cite{time1}-\cite{null}. Among others, solutions
with the linear dilaton background in the null direction with
$\frac12$ supersymmetry are proposed to describe the singular
region in the spacetime without difficulty~\cite{bb1}, and various
extensions have been considered~\cite{bb2}-\cite{NPS}. For a
detailed review of the big-bang models in string theory see
\cite{BC}. It has been argued that it is possible to map the
theory near the singularity at a very early time to the dual
matrix theory which allows us to discuss the behaviors of the
solution in perturbative picture, whereas in the far future it can
be thought of as a perturbative string theory with weak string
coupling. So it is the matrix degrees of freedom but not the
perturbative string states that describe the physics near the
singularities, but later the spacetime picture in terms of the
closed strings becomes relevant. Furthermore the celebrated
AdS/CFT duality has also been used to argue that the dual field
theory is a time-dependent supersymmetric gauge theory on the
boundary of the AdS space in other backgrounds. In view of these
interesting developments in the study of time-dependent solutions
in string theory, it is important to explore further examples of
supersymmetric solutions of string and supergravity theories in
time-dependent backgrounds.

On the other hand, D-branes can probe the nonperturbative dynamics of the string
theory and they have been used to study various duality aspects of
string theory. A systematic derivation of the general D-brane
solutions in the pp-wave backgrounds has been given in \cite{OPS}.
It is thus interesting to find if we can have such brane solutions
in time-dependent backgrounds with time-dependent dilaton. In
fact, D3-brane solutions have been found and discussed in
\cite{CH,DMNT} and other single brane solutions in \cite{NP,NPS},
but it is not known if there exist further general intersecting
brane solutions of this type in a time-dependent set up.

Motivated by the recent surge of interest in finding out time-dependent
solutions in supergravity and speculation on the dual field theory
in this set up, in this paper we present a general class of
intersecting brane solutions in time-dependent supersymmetric
backgrounds with and without pp-wave by using the method developed
in Ref.~\cite{NO,OPS}. We start with a general ansatz for the
metric and solve for the field equations of the supergravity. We
also derive the intersection rules for the branes in this
background. Some examples of dual field theories are also
discussed briefly.

The rest of the paper is organized as follows. In sect.~2 we
present the classical solution of supergravity by  directly
solving the equations of motion. We, for simplicity, take the
factorized ansatz for the metric functions as the product of usual
$r$-dependent part and the time-dependent part. This way of taking
the ansatz has a clear advantage that the supergravity equations
of motion essentially have two different contributions coming from
the $r$-dependent part and a differential equation involving only
time derivatives. We then present three kinds of solutions. We
further show that special cases of our general solutions reproduce
known solutions. We also write explicitly a particular
example of intersecting branes of D1-D5 system with a linear
dilaton and make some comments about the dual field theory. In
sect.~3 we present our conclusions and outlook.

\section{Supergravity solution}

The low-energy effective action for the supergravity system
coupled to dilaton and $n_A$- form field strength is given by
\bea
I = \frac{1}{16 \pi G_D} \int d^D x \sqrt{-g} \left[
 R - \frac12 (\pa \phi)^2 - \sum_{A=1}^m \frac{1}{2 n_A!} e^{a_A \phi}
 F_{n_A}^2 \right],
\label{action}
\ena
where $G_D$ is the Newton constant in $D$
dimensions and $g$ is the determinant of the metric. The last term
includes both RR and NS-NS field strengths and $a_A = \frac12
(5-n_A)$ for RR field strengths and $a_A = -1$ for NS-NS 3-form.
We put fermions and other background fields to be zero.

{}From the action (\ref{action}), one can derive the field
equations/Bianchi identities
\bea
R_{\mu\nu} = \frac12 \pa_\mu
\phi \pa_\nu \phi + \sum_{A} \frac{1}{2 n_A!}
 e^{a_A \phi} \left[ n_A \left( F_{n_A}^2 \right)_{\mu\nu}
 - \frac{n_A -1}{D-2} F_{n_A}^2 g_{\mu\nu} \right],
\label{Einstein}
\ena
\bea
\Box \phi = \sum_{A} \frac{a_A}{2 n_A!}
e^{a_A \phi} F_{n_A}^2,
\label{dila}
\ena
\bea
\pa_{\mu_1} \left(
\sqrt{- g} e^{a_A \phi} F^{\mu_1 \cdots \mu_{n_A}} \right) = 0,
\label{field}
\ena
\bea
\pa _{[\mu} F_{\mu_1 \cdots \mu_{n_A}]} =0.
\label{bianchi}
\ena

In this paper we consider the case in which the metric functions
depend on $u$ and $r$. Specifically we take the metric ansatz \bea
ds_D^2 = e^{2(u_0+v_0)} [-2dudv + K(u,y_\a, z_i) du^2] +
\sum_{\a=1}^{d-2} e^{2 (u_\a+v_\a)} dy_\a^2 +e^{2(B+C)}[dr^2 + r^2
d\Omega_{\tilde d +1}^2], \label{met} \ena where $D=d+\tilde d+2$,
the coordinates $u$, $v$ and $y_\a, (\a=1,\ldots, d-2)$
parameterize the $d$-dimensional world-volume directions and the
remaining $\tilde d + 2 $ coordinates $r$ and angles are
transverse to the brane world-volume (sometimes denoted also by
the orthogonal coordinates $z_i$), $d\Omega_{\tilde d+1}^2$ is the
line element of the $(\tilde d+1)$-dimensional sphere. Here $u_0,
u_\a$ and $B$ are assumed to be functions of $r$ only, and $v_0,
v_\a$ and $C$ are those of $u$ only. The dilaton $\phi$ is also
taken as a sum of $u$-dependent and $r$-dependent terms: $\phi =
\phi_r + \phi_u$. Our ansatz includes more general solutions than
those in \cite{CH,DMNT}.

For the field strength backgrounds, we take
\bea
F_{n_A} = e^{2f_A(u)} E_A'(r) du \wedge dv \wedge dy^{\a_1} \wedge \cdots \wedge
dy^{\a_{q_A -1}} \wedge dr,
\label{eleb}
\ena
where $n_A = q_A +2$.
Throughout this paper, the dot and prime denote derivatives with
respect to $u$ and $r$, respectively. Eq.~\p{eleb} is an electric
background and we could also consider magnetic background, but
that is basically the same as the electric case with the
replacement
\bea
g_{\mu\nu} \to g_{\mu\nu}, \quad F_{n} \to e^{a
\phi}*\! F_{n}, \quad \phi \to - \phi. \label{sdual}
\ena
This is due to the S-duality symmetry of the original system~\p{action}.
So we do not have to consider it separately.

The NS-NS 3-form responsible for the off-diagonal component of the
metric is separately written as
\bea
H_{uij} = e^{2g(u)} \pa_{[i} b_{j]},
\label{nsns}
\ena
such that it satisfies the Bianchi identity. Here the indices $i,j$
denote the directions transverse to the branes ($z_i, z_j$ or $r$ and angles).
The field equation for this NS-NS 3-form leads to
\bea
\pa_{[i} b_{j]} = e^{-U+2(u_0+v_0+B+C)+\phi}\mu_{ij},
\label{ns}
\ena
where $\mu_{ij}$ is constant and $U$ is defined by
\bea
U = 2u_0+ \sum_{\a=1}^{d-2} u_\a +\tilde d B.
\ena
There is no restriction on the $u$-dependent part of $b_j$ from the field equation,
but we have chosen it like eq.~\p{ns} as our convention.

With our ansatz, the Einstein equations~\p{Einstein} reduce to
\bea
&& u_0'' + u_0'\Big\{ U' + \frac{\tilde d+1}{r} \Big\}
= \sum_A \frac{D-q_A-3}{2(D-2)} S_A T_A(E'_A)^2,
\label{fe1} \\
&& \sum_{\a=1}^{d-2} \ddot v_\a + (\tilde d + 2) \ddot C +
\sum_{\a=1}^{d-2} {\dot v_\a}^2 + (\tilde d + 2) {\dot C}^2 - 2
\dot v_0 \{ \sum_{\a=1}^{d-2} \dot v_\a + (\tilde d + 2) \dot C\} \nn
&& +\; e^{2(u_0+v_0 - B-C)}K \Bigg[ u''_0 + \frac{\tilde d+1}{r}u'_0
+ \frac12 K^{-1}\Box^{(\tilde d + 2)} K \nn
&& \hs{35}
+\; \pa_i \Big(u_0+\frac12 \ln K\Big) \pa^i U
\Bigg] + \frac12 \sum_{\a=1}^{d-2} e^{2(u_0+v_0-u_\a-v_\a)}\pa^2_\a K \nn
&& = -\sum_A\frac{D-q_A-3}{2(D-2)} e^{2(u_0+v_0-B-C)} K S_A T_A(E'_A)^2 \nn
&& \hs{30} +\; \frac{1}{4} e^{-2U-(4B +4C+\phi) + 4g(u)}(\pa_{[i} b_{j]})^2
- \frac12 (\dot \phi)^2,
\label{fe2}
\\
&& u_\a'' + u_\a'\Big( U' + \frac{\tilde d+1}{r} \Big)
= \sum_A \frac{\d^{(\a)}_A}{2(D-2)} S_A T_A(E'_A)^2,
\label{fe3}
\\
&& U''+ B'' - B'\Big( 2u_0' + \sum_{\a=1}^{d-2} u_\a' - \frac{\tilde d+1}{r}\Big)
+2(u_0')^2 + \sum_{\a=1}^{d-2} (u_\a')^2 \nn
&& = - \frac12 (\phi')^2 + \sum_A\frac{D-q_A-3}{2(D-2)}S_A T_A(E'_A)^2,
\label{fe4}
\\
&& B'' -\frac{1}{r^2}+\Big(B' + \frac{1}{r}\Big) \Big( U'
+ \frac{\tilde d+1}{r} \Big) - \frac{\tilde d}{r^2}
= - \sum_A\frac{q_A+1}{2(D-2)}S_A T_A(E'_A)^2,
\label{fe5}
\ena
where $S_A, T_A$ and $\d_A^{(\a)}$ are defined by
\bea
&& S_A = \exp\Big({\e_A a_A\phi_r - 2\sum_{\a \in q_A} u_\a}\Big),\quad
T_A = \exp\Big({\e_A a_A\phi_u - 2\sum_{\a \in q_A} v_\a + 4f_A}\Big),
\label{sa}
\ena
\bea
\d_A^{(\a)} = \left\{
\begin{array}{l}
D-q_A-3 \\
-(q_A+1)
\end{array}
\right. \hs{5} {\rm for} \hs{3} \left\{
\begin{array}{l}
y_\a \mbox{   belonging  to $q_A$-brane} \\
{\rm otherwise}
\end{array}
\right. ,
\ena
respectively, the sum of $\a$ runs over the world-volume of the $q_A$-brane
($u,v$ and $(q_A-1)\; y^\a$ coordinates, and so
$\sum_{\a \in q_A} u_\a =2u_0 + \sum_{\a=1}^{q_A-1} u_\a$ for example),
and $\e_A= +1 (-1)$ is for electric (magnetic) backgrounds. The
equations (\ref{fe1}), (\ref{fe2}), (\ref{fe3}), (\ref{fe4}) and
(\ref{fe5}) are the $uv, uu, \a\b, rr$ and $ab$ components of the
Einstein equations (\ref{Einstein}), respectively.
The dilaton eq.~\p{dila} and the remaining eqs.~\p{field} and/or \p{bianchi}
yield
\bea
e^{-U} ( e^U \phi')' &=& -\frac12 \sum_A\e_A a_A S_A T_A (E'_A)^2,
\label{dil} \\
\Big( r^{\tilde d+1} S_A E_A' \Big)' &=& 0,
\label{fe6}
\ena
where the dilaton field is written as a sum of $r$- and
$u$-dependent terms: $\phi=\phi_r + \phi_u$.

The field eqs.~\p{fe1} -- \p{fe5}, \p{dil} and \p{fe6} are simplified
considerably by imposing the condition
\bea
U = 0.
\label{conde}
\ena
It is known that under this condition, all the supersymmetric intersecting
brane solutions can be derived~\cite{NO}.

Now the dilaton eq.~(\ref{dil}) gives
\bea
\phi'' + \frac{(\tilde d+1)}{r}\phi'  = -\frac12 \sum_A\e_A a_A S_A T_A (E'_A)^2,
\label{dile}
\ena
In order to cancel the $u$-dependence in eq.~\p{dile},
we should set
\bea
4f_A = - \e_A a_A \phi_u +2 \sum_{\a\in q_A} v_\a,
\ena
namely $T_A=1$. Then the field equation for the dilaton reduces to
the one discussed in \cite{OPS}.

Thus, for our ansatz that the exponents of metric functions, the
dilaton and other backgrounds are simply sums of terms dependent
on $u$ and $r$, we see that the Einstein equations and other field
equations separate into terms dependent on $u$ and $r$. The metric
corresponds to the ``warped form'' of supersymmetric
time-dependent solutions and static branes, giving a
generalization of D3-branes in time-dependent backgrounds in
Refs.~\cite{CH,DMNT}. So these parts in the field equations should
be separately satisfied. The $r$-dependent equations are simply those of
Ref.~\cite{OPS} for static branes in the pp-wave backgrounds
except for eq.~\p{fe2}. These $r$-dependent equations are already solved
in Ref.~\cite{OPS}.

For eq.~\p{fe2}, the $u$-dependent part should balance with each other:
\bea
- \sum_{\a=1}^{d-2} \ddot v_\a - (\tilde d + 2) \ddot C
-\sum_{\a=1}^{d-2} {\dot v_\a}^2 - (\tilde d + 2) {\dot C}^2 + 2
\dot v_0 \Big\{ \sum_{\a=1}^{d-2} \dot v_\a + (\tilde d + 2) \dot C \Big\}
= \frac12 (\dot \phi_u)^2 ,
\label{phi_u}
\ena
which is the only constraint that the $u$-dependent terms should satisfy.

Here we note that there are three classes of solutions depending on whether
we have the plane-wave function $K$ or not and which coordinate dependence it has.
We now discuss these solutions separately.

\subsection{Brane solutions in time-dependent backgrounds}
\label{c1}

If we do not have the function $K$, its source should be zero:
\bea
b_i=0,
\ena
which is enough to eliminate the $u$-dependence from eq.~\p{fe2},
and the only condition that $v_\a$ should satisfy is eq.~\p{phi_u}.
The resulting equations for $r$-dependent factors are the same as
the static branes~\cite{NO} and we find the solutions are simply given by
\bea
ds_D^2 &=& \prod_A H_A^{2 \frac{q_A+1}{\Delta_A}}
\Bigg[ - e^{2v_0(u)} \prod_A H_A^{- 2 \frac{D-2}{\Delta_A}} 2dudv \nn
&& \hs{20} + \; \sum_{\a=1}^{d-2} \prod_A H_A^{- 2 \frac{\c_A^{(\a)}}{\Delta_A}}
e^{2v_\a(u)} dy_\a^2 + e^{2C(u)}(dr^2 + r^2 d\Omega_{\tilde d+1}^2) \Bigg], \nn
&& E_A = \sqrt{\frac{2(D-2)}{ \Delta_A}} H^{-1}_A, \quad
\phi = \sum_{A} \epsilon_A a_A \frac{D-2}{\Delta_A} \ln H_A +
\phi_u,
\ena
where $\Delta_A$ and $\c_A^{(\a)}$ are defined by
\bea
&& \Delta_A = (q_A + 1) (D-q_A-3) + \frac12 a_A^2 (D-2), \nn
&& \c_A^{(\a)} = \left\{ \begin{array}{l}
D-2 \\
0
\end{array}
\right. \hs{5} {\rm for} \hs{3} \left\{
\begin{array}{l}
y_\a \mbox{   belonging  to $q_A$-brane} \\
{\rm otherwise}
\end{array},
\right.
\label{gamma}
\ena
respectively, and $\phi_u$ should be determined by the relation~\p{phi_u}.
Here and in what follows, $H_A=1+\frac{Q_A}{r^{\tilde d}}$ represent harmonic
functions in $\tilde d+2$ dimensions.
These give the generalization of the class of solutions discussed
in Ref.~\cite{DMNT}, in which D3-brane solutions are given with $C=0$.
Our solutions generalize these to orthogonally intersecting branes
with nonvanishing $C$.

\subsection{Branes in plane-wave backgrounds}
\label{c2}

If $K$ does not depend on $y_\a$, $\pa_\a^2 K$ term in eq.~\p{fe2} is absent.
Considering eq.~\p{ns}, we find that the condition
\bea
4g(u) = 2(v_0+C)+\phi_u,
\label{cond1}
\ena
is enough to eliminate $u$-dependence from eq.~\p{fe2}. It is then easy to
give the solutions, following Ref.~\cite{OPS}.

The result is
\bea
ds_D^2 &=& \prod_A H_A^{2 \frac{q_A+1}{\Delta_A}}
\Bigg[ e^{2v_0(u)} \prod_A H_A^{- 2 \frac{D-2}{\Delta_A}} \Big\{ -
2dudv + K du^2\Big\} \nn
&& \hs{30} + \;
\sum_{\a=1}^{d-2} \prod_A H_A^{- 2 \frac{\c_A^{(\a)}}{\Delta_A}}e^{2v_\a}
dy_\a^2 + e^{2C(u)}(dr^2 + r^2 d\Omega_{\tilde d+1}^2) \Bigg], \nn
&& E_A = \sqrt{\frac{2(D-2)}{ \Delta_A}} H^{-1}_A, \quad
\phi = \sum_{A} \epsilon_A a_A \frac{D-2}{\Delta_A} \ln H_A +
\phi_u,
\ena
where $\c_A^{(\a)}$ is the same as in eq.~\p{gamma} and the function $K$
is defined by
\bea
\Box^{(\tilde d + 2)} K = - \frac12 (\mu_{ij})^2 \prod_A H_A^{l_A},
\label{ka0}
\ena
and $\phi_u$ should be determined by the relation~\p{phi_u}.

In Ref.~\cite{DMNT}, D3-brane solutions with $\mu_{ij}=0$ and $C=0$
are given. There $K$ is taken to depend only on $u$, in which case
we see that eq.~\p{ka0} is trivially satisfied and $K$ can be an arbitrary
function of $u$, in agreement with Ref.~\cite{DMNT}. The same solutions
in different coordinate system are also given in~\cite{CH}. Our solutions,
when restricted to single branes, are still more general than those.

\subsection{Branes in more general wave backgrounds}
\label{c3}

If we have the general function $K$ describing waves, we must have
\bea
4g(u) = 2(v_0+C)+\phi_u, \quad
v_\a = C,
\label{cond2}
\ena
in order to get rid of $u$-dependence from eq.~\p{fe2}.
Then we see that eq.~\p{phi_u} reduces to
\bea
-(D-2) ( \ddot C + \dot C^2 -2 \dot v_0 \dot C) = \frac12 (\dot \phi_u)^2.
\ena
and the remaining condition from eq.~\p{fe2} precisely gives
the corresponding one in Ref.~\cite{OPS}.

The determination of the remaining functions are essentially
the same as \cite{OPS}, so we do not repeat the details,
but simply present the final result:
\bea
ds_D^2 &=& \prod_A H_A^{2 \frac{q_A+1}{\Delta_A}}
\Bigg[ e^{2v_0(u)} \prod_A H_A^{- 2 \frac{D-2}{\Delta_A}} \Big\{ -
2dudv + K du^2\Big\} \nn
&& \hs{30} + \; e^{2C(u)}
\sum_{\a=1}^{d-2} \prod_A H_A^{- 2 \frac{\c_A^{(\a)}}{\Delta_A}}
dy_\a^2 + e^{2C(u)}(dr^2 + r^2 d\Omega_{\tilde d+1}^2) \Bigg], \nn
&& E_A = \sqrt{\frac{2(D-2)}{ \Delta_A}} H^{-1}_A, \quad
\phi = \sum_{A} \epsilon_A a_A \frac{D-2}{\Delta_A} \ln H_A +
\phi_u,
\ena
where $\c_A^{(\a)}$ is the same as in eq.~\p{gamma} and the function $K$
is defined by
\bea
\Big(\Box^{(\tilde d + 2)}  +
\sum_{\a=1}^{d-2} \prod_A H_A^{2\frac{\c_A^{(\a)}}{\Delta_A}}
\pa^2_\a \Big) K = - \frac12 (\mu_{ij})^2 \prod_A H_A^{l_A},
\label{ka1}
\ena
respectively, and $\phi_u$ should be determined by the relation~\p{phi_u}.
The function $K$ is essentially the same as given in Ref.~\cite{OPS}.
For a single D$q_A$-brane, eq.~(\ref{ka1}) admits a solution of the form
\bea
K = c + \frac{\mathcal Q}{r^{\tilde d}} -\frac{(\mu_{ij})^2}{32} \Big(r^2
+ \sum_{\a}y_\a^2 + \frac{(q_A-1)}{(\tilde d-2)}\frac{Q_A}
{r^{\tilde d -2}}\Big),
  ~~~({\rm for}~~\tilde d \ne 2)
\label{ka2}
\ena
and
\bea
K = c + \frac{\mathcal Q}{r^{\tilde d}}
-\frac{(\mu_{ij})^2}{32} \Big(r^2 + \sum_{\a}y_\a^2 -
(q_A-1)Q_A\ln r\Big),
 ~~~({\rm for }~~ \tilde d = 2)
\label{ka3}
\ena

We also have the intersection rules for the branes~\cite{OPS}. If
$q_A$-brane and $q_B$-brane intersect over ${\bar q} (\leq q_A, q_B)$
dimensions, this gives
\bea
{\bar q} = \frac{(q_A+1)(q_B+1)}{D-2}-1 - \frac12 \e_A a_A \e_B a_B.
\label{ints}
\ena
For D-branes
$\e_A a_A = \frac{3-q_A}{2},$
and we get
\bea
\bar q=\frac{q_A+q_B}{2} -2. \label{drule}
\ena
The results presented here are the generalization of the intersection
rules already discussed in the literature~\cite{NO} to the supersymmetric
intersecting branes in time-dependent backgrounds.

\subsection{Supersymmetry}

In the time-dependent background without the NS-NS flux,
the amount of unbroken supersymmetry depends on the condition
$\gamma_u \e = 0$, and the brane supersymmetry condition.
Usually these two conditions are two independent conditions,
and one needs to impose them both on the gravitino and the
dilatino variations. They are compatible with each other,
but the unbroken supersymmetry is 1/2 compared with the usual static
brane solutions.
So if $\mu_{ij}=0$, we have 1/2 supersymmetry
for no branes, 1/4 supersymmetry for single branes as discussed in
Refs.~\cite{CH,DMNT}, 1/8 supersymmetry for 2 orthogonally
intersecting branes and so on.

In the presence of the NS-NS flux $H_{uij}$,
the supersymmetry variations normally restrict the form
of the flux for getting a solution of the killing spinor equation
of motion. As it was argued in \cite{OPS}, this choice further
breaks 1/2 of the remaining supersymmetry.

To get an idea how the solutions look like, let us consider
intersecting D1-D5 system without $\mu_{ij}$ (the case in subsect.~2.2).
The supergravity solution is given by
\bea
ds^2 &=& H^{-\frac{3}{4}}_1
H^{-\frac{1}{4}}_5 e^{2v_0(u)}\left(-2 du dv + K(u) du^2\right) +
\left(\frac{H_1}{H_5}\right)^{\frac{1}{4}}
\sum^4_{\alpha =1} e^{v_\a(u)} dy^2_\alpha \nn
&& + \; H^{\frac{1}{4}}_1 H^{\frac{3}{4}}_5
e^{2C(u)}\left(dr^2 + r^2 d\Omega^2_3\right), \nn \phi
&=& \ln \left(\frac{H_1}{H_5}\right)^{\frac{1}{2}} + \phi_u.
\label{d1-d5}
\ena
We see that the solution resembles the ordinary intersecting branes,
and the difference is the presence of time-dependent factors multiplying
the metric and terms appearing in the dilaton and forms such as \p{eleb}.
In the present intersecting branes in a wave background
with $\mu_{ij}=0$, the amount of unbroken supersymmetry is four,
as this preserves 1/8 of the full type IIB supersymmetry,
corresponding to $N=2$ supersymmetry in 2 dimensions.
If the metric depending on the light-cone coordinate $u$ are finite,
the near horizon geometry is $AdS_3 \times S^3 \times M^4$.
As argued in Ref.~\cite{MS}, the corresponding dual field theory would be
two-dimensional conformal field theory, now in time-dependent backgrounds.

The D1-D5 system has a singularity (big bang) at
$v_0(u)\rightarrow -\infty$, as some components of the metric
vanish and we can easily see that the curvature components also
blow up. For example, let us choose the dilaton and metric
functions linear in the coordinate $u$: $\phi=-au\; (a>0),
v_0(u)=v_\a(u)=bu, C(u)=cu$. The condition~\p{phi_u} tells us that
$a^2+16c^2=8(b+c)^2.$
For the simple choice $c=K=0$ and $b=a/2\sqrt{2}$, there appears
a singularity in the infinite past in the solution~(\ref{d1-d5}).

\section{Concluding remarks}

Motivated by the recent interest in supersymmetric time-dependent
solutions in supergravity with their possible application to the
singularities in our spacetime, we have derived a rather general
class of solutions with and without pp-wave. Our solutions reproduce
many of the known time-dependent solutions but are more
general than those already known. These solutions have
time-dependent null dilaton which is related to the string coupling.
So the string theory has time-dependent coupling.

According to the AdS/CFT correspondence, it is expected that these
solutions have dual description in terms of super Yang-Mills
theories. It would be very interesting to further explore the
nature of these solutions, especially their singularity
structures, and their field theory duals. An interesting question
is whether and how the field theory dual gives well-defined
description of the behavior of the solutions close to the
singularities. It would also be important to find more
cosmological applications of these solutions and try to understand the
nature of the space-like singularities.

\vspace{5mm}

\section*{Acknowledgements}

The work of NO was supported in part by the Grant-in-Aid for
Scientific Research Fund of the JSPS No. 16540250.


\end{document}